\documentclass[]{aa} 
\usepackage{natbib}
\bibpunct{(}{)}{;}{a}{}{,} 
\usepackage{graphicx}
\usepackage{longtable}
\usepackage{lscape}
\usepackage{txfonts}
\usepackage{color}

\begin{document}

   \title{The visible and near-infrared spectra of asteroids in cometary orbits}


   \author{J. Licandro\inst{1,2}
          \and
          M. Popescu\inst{1,2,3}
          \and
         J. de Le\'on\inst{1,2}
          \and
          D. Morate\inst{1,2}
          \and
          O. Vaduvescu\inst{4,1,2}
          \and 
          M. De Pr\'a\inst{5}
          \and
          V. Al\'{i}-Lagoa\inst{6}
           }

   \institute{Instituto de Astrof\'{\i}sica de Canarias (IAC), C\/V\'{\i}a L\'{a}ctea s\/n, 38205 La Laguna, Spain
              \and
              Departamento de Astrof\'{\i}sica, Universidad de La Laguna, 38206 La Laguna, Tenerife, Spain
              \and
              Astronomical Institute of the Romanian Academy, 5 Cu\c{t}itul de Argint, 040557 Bucharest, Romania
               \and
               Isaac Newton Group of Telescopes, Apto. 321, E-38700 Santa Cruz de la Palma, Canary Islands, Spain
               \and
               Departamento de Astrof\'{\i}sica, Observat\'orio Nacional, Rio de Janeiro, 20921-400, Brazil
               \and
               Max-Planck-Institut fur extraterrestrische Physik (MPE), Giessenbachstrasse 1, 85748 Garching, Germany
                }
     
   \date{Received XXX xx, XXX; accepted XXX xx, XXX}

 
  \abstract
  {Dynamical and  albedo properties suggest that asteroids in cometary orbits (ACOs)  are dormant or extinct comets. Their study provides new insights for understanding the end-states of comets and the size of the comet population.}
  {We intend to study the visible and near-infrared (NIR) spectral properties of different ACO populations and compare them to the independently determined properties of comets. }
   {We select our ACOs sample based on published dynamical criteria and present our own observational results obtained using the  10.4m Gran Telescopio Canarias (GTC), the 4.2m William Herschel Telescope (WHT), the 3.56m Telescopio Nazionale Galileo (TNG), and the 2.5m Isaac Newton Telescope (INT), all located at the El Roque de los Muchachos Observatory (La Palma, Spain), and the 3.0m  NASA Infrared Telescope Facility (IRTF), located at the Mauna Kea Observatory, in Hawaii.  We include in the analysis the spectra of ACOs obtained from the literature. We derive the spectral class and the visible and NIR spectral slopes. We also study the presence of hydrated minerals by studying the 0.7 $\mu$m band and the UV-drop below 0.5 $\mu$m associated with phyllosilicates. }
   {We present new observations of 17 ACOs, 11 of them observed in the visible, 2 in the NIR and 4 in the visible and NIR. We also discuss the spectra of 12 ACOs obtained from the literature. 
 All but two ACOs have a primitive-like class spectrum (X or D-type).  Almost 100\% of the ACOs in long-period cometary orbits (Damocloids) are D-types. Those in Jupiter family comet orbits (JFC-ACOs) are $\sim$ 60\% D-types and $\sim$ 40\% X-types. The mean spectral slope $S'$  of JFC-ACOs is 9.7 $\pm$ 4.6 \%/1000 \AA \   and for the Damocloids this is 12.2 $\pm$ 2.0 \%/1000 \AA . No evidence of hydration on the surface of ACOs is found from their visible
spectra. The spectral slope and spectral class distribution of ACOs is similar to that of comets.}
   {The spectral taxonomical classification and the spectral slope distribution of ACOs, and the lack of spectral features indicative of the presence of hydrated minerals on their surface, strongly suggest that ACOs are likely dormant or extinct comets. }
   


   \titlerunning{The visible and near-infrared spectra of ACOs}
   
   \maketitle
%

\section{Introduction}

Asteroids in cometary orbits (ACOs) are objects in typical cometary orbits that have never shown any kind of activity. Several classification schemes have been used to identify  ACOs, all of them based on Tisserand's  parameter ($T_J$), a constant of motion in the restricted three-body problem related to an object's encounter velocity with Jupiter. The orbits of the large majority of asteroids present $T_J>3$, while most comets have orbits with $T_J<3$. \citet{Fernandez05} introduced a second criterion using the minimum orbital intersection distance (MOID) with Jupiter. To avoid objects with stable dynamical evolution incompatible with the chaotic dynamics of comets, \citet{Tancredi2014} presented a criterion to identify ACOs that ensures the selection of objects with a dynamical evolution similar to the population of periodic comets. Tancredi also produced a list of 331 ACOs based on the proposed classification criterion applied to a sample of $>$500,000 known asteroids that, from a dynamical point of view, are the best known extinct/dormant comet candidates. ACOs in Tancredi's list are classified in subclasses similar to the cometary classification: 203 objects belong to the Jupiter family group (JFC-ACOs); 72 objects are classified as Centaurs; and 56 objects have Halley-type or long period comet (LPC)-type orbits (also known as ``Damocloids'', \citealt{Jewitt2005}). The classical classification of comets, depending on whether their orbital periods are less than or greater than 200 yr, provides a good discriminant between comets originated in the trans-Neptunian belt (JFCs) or the Oort Cloud (LPCs). Therefore, JFC-ACOs and Damocloids should also be scattered objects from the trans-Neptunian belt and the Oort Cloud, respectively.

Observations of other physical properties (albedo, spectra, rotation period, etc.) of the ACOs in the Tancredi's list are needed to determine whether these objects truly are extinct/dormant comets or simply asteroids that have escaped from the main belt into cometary-like chaotic orbits.  Recently, \citet{licandro16} determined the geometric albedo ($p_V$), beaming parameter ($\eta$) and size distribution of a significant number of JFC-ACOs and Damocloids from Tancredi's list using data from NASA's Wide-field Infrared Explorer (WISE), and concluded that the $p_V$- and $\eta$-value distributions of ACOs are very similar to those of JFCs nuclei.

Spectral properties of ACOs have also been explored, including spectrophotometric and spectroscopic studies in the visible and near-infrared (NIR), and compared to the spectral properties of cometary nuclei  (\citealt{licandro2006,licandro08,DeMeo2008,AlvarezCandal2013,Jewitt2005,Jewitt2015b}). These studies feature a significant number of asteroids that do not satisfy Tancredi's criterion, and therefore include several objects that are unlikely comets.  Visible and NIR spectral properties of 39 ACOs are presented in \citet{licandro08}, but only 6 of them are in Tancredi's list. Another 6 objects from the 55 ACOs studied by \citet{DeMeo2008} are included in this list. On the other hand, the large majority of Damocloids in \cite{Jewitt2005,Jewitt2015b} are in Tancredi's list. Therefore, the sample of ACOs in \citet{Tancredi2014}, in particular the JFC-ACOs, with known spectral properties, is small ($\sim$ 4\% of the sample). This sample needs to be enlarged.
 
The aim of this paper is to study the visible and NIR spectral properties of ACOs selected using Tancredi's criteria, compare them with  the spectral properties of comets and validate Tancredi's hypothesis: that these are the best dormant comet candidates. We present new spectra of 17 ACOs, 11 of them observed in the visible, 2 in the NIR and 4 in the visible and NIR, and analyze the spectral properties of the ACO population using also the spectra of 12 ACOs obtained from the literature. 

The paper is organized as follows. In Sect. \ref{sec:obs},  observations, data reduction, and spectral extraction are described. In Sect. \ref{sec:literature}, the spectra of ACOs retrieved from the literature are presented. In Sect. \ref{sec:results}, we analyze the spectral properties of the ACO population (taxonomical and spectral slope distribution, and hydration signatures) using the spectra presented in the preceding sections. We present the results and analysis considering two ACO populations: those in Jupiter family comet orbits  (JFC-ACOs) and those in Halley-type orbits (Damocloids), assuming they likely have an origin in two different comet reservoirs: the trans-Neptunian belt and the Oort cloud, respectively. In Sect. \ref{sec:comparison}, the spectral properties of ACOs are compared with the known spectral properties of cometary nuclei.  Finally, the conclusions are presented in Sect. \ref{sec:conclusions}.

\section{Observations \label{sec:obs}}

Visible and/or NIR spectra of 17 ACOs were obtained during different nights using four different telescopes at the ``El Roque de los Muchachos Observatory" (ORM, Canary Islands, Spain): the 10.4 GTC,  the 4.2m WHT, the 3.56m TNG, and the  2.5m INT. Additionally, we  used the 3m IRTF located in Mauna Kea Observatory (Hawaii, USA). The list of observed objects and their main orbital elements are shown in Table \ref{table:elements}. The majority of the observed ACOs (12) are JFC-ACOs while the other 5 are Damocloids.  Only two of the observed objects are not in Tancredi's list. These ACOs are Damocloids, and are not in Tancredi's list because the uncertainty of their orbits prevents them from fulfilling Tancredi's criteria. However, we included them, because despite their orbital uncertainties, both objects are still Damocloids. Between the observed ACOs,  2006 HR$_{30}$ was later reported active \citep{Hicks2007} and is actually classified as comet P/2006 HR$_{30}$ (Siding Spring).

\begin{table*}
\caption{\label{table:elements}  Orbital elements of the observed ACOs taken from the NASA JPL Horizons ephemeris service (http://ssd.jpl.nasa.gov/horizons.cgi). {\it Group} is the dynamical subclass of the object}
\vskip4mm
\centering
\begin{tabular}{lcrcrc}     
\hline
ACO & Group & $a$ (au) & $e$ & $i$ ($^{\circ}$)& $T_J$~\\ \hline
(6144) Kondojiro                & JFC-ACO & 4.757 &0.362 & 5.88 & 2.867 \\
(18916) 2000 OG$_{44}$  & JFC-ACO & 3.851 &0.585 & 7.41 & 2.735 \\
(30512) 2001 HO$_{8}$           & JFC-ACO & 3.842 &0.324 & 25.77 & 2.819  \\
(248590) 2006 CS                & JFC-ACO & 2.911 &0.698 & 52.36 & 2.441 \\
(315898) 2008 QD$_{4}$          & JFC-ACO & 8.395 &0.350 & 42.02 & 2.388  \\ 
(347449) 2012 TW$_{236}$                & JFC-ACO & 6.986 &0.571 & 11.90 & 2.607 \\
(366186)        2012 HE$_{26}$  & JFC-ACO & 3.586 &0.303 & 14.18 & 2.985 \\
(380282)        2002 AO$_{148}$         & JFC-ACO & 5.72 &0.267 & 19.12 & 2.819  \\
(397262)        2006 RN$_{16}$  & JFC-ACO & 4.447 &0.398 & 12.79 & 2.824  \\
(405058) 2001 TX$_{16}$         & JFC-ACO & 3.579 &0.598 & 8.14 & 2.770  \\
(465293)  2007 TR$_{330} $      & JFC-ACO & 3.621 &0.404 & 11.40 & 2.933  \\
2002 RP$_{120}$ & Damocloid & 54.377 &0.955 & 118.91 & -0.834  \\
2005 NA$_{56}$  & JFC-ACO & 3.029 &0.544 & 9.14 & 2.982  \\
P/2006 HR$_{30}$ (Siding Spring)        & Damocloid & 7.818 &0.843 & 31.88 & 1.785  \\
2006 EX$_{52}$  & Damocloid & 42.774 &0.940 & 150.16 & 1.579 \\
2007 DA$_{61}   $ & Damocloid & 495.167 &0.994 & 76.77 & 0.474 \\
2009 FW$_{23}   $ & Damocloid & 11.517 &0.855 & 86.66 & 0.542 \\
\hline
\end{tabular}
\end{table*}

\begin{table*}
\caption{\label{table1} Observational circumstances of the ACOs presented in this work. Information shown in this table includes the airmass (X), heliocentric ($r$) and geocentric ($\Delta$) distances and phase angle ($\alpha $) at the time of the observation.}

\vskip4mm
\centering
\begin{tabular}{rccccrccrrrc}     
\hline
ACO & Group & Telescope & Date & UT start & X & $r$ & $\Delta$ & $\alpha $ &  \#$^1$ & Exp.$^2$ & Solar\\
        &               &       &       &               &       &(au)   &(au)   &($^{\circ}$)   &       &(sec)  &Analogue\\
            
\hline            
(6144)          & JFC-ACO & INT & 07/02/2015 &05:46 &1.5 &1.512  &3.036 &3.0  & 4 & 700 &HD142801\\
(6144)          & JFC-ACO & IRTF &14/05/2015  &11:31 &1.2 &3.116 &2.120 &3.8 &16 &120 &HD1444821\\
(18916)                 & JFC-ACO & INT & 22/07/2015 &03:48 &1.5 &2.494  &1.519 &8.4  &3  & 900 &HD198273\\
(30512)                 & JFC-ACO & GTC & 17/12/2014 &04:23 & 1.1 & 2.997 & 2.167 & 11.9 & 3 & 250 &SA 98-978 \\
(30512)                 & JFC-ACO & INT & 02/03/2015   &21:32 &1.0 &2.789  &2.186 &18.4 &3 &700 &HD259516\\
(30512)                 & JFC-ACO & IRTF & 07/02/2015 &08:01 &1.0 &2.864 &1.947 &8.8 &32 &120 &HD56513\\
(248590)                & JFC-ACO & IRTF & 09/04/2016 &05:44 &1.4 &0.885 &0.480 &89.3 &6 &120 &HD29714\\
(315898)                & JFC-ACO & GTC & 19/01/2015 &05:11 & 1.1 & 8.161 & 7.465 & 5.1 & 3 & 500 &SA 102-1081\\
(347449)                & JFC-ACO & INT & 30/12/2014 &02:42 &1.1 &3.290  &2.347 &5.8   &3 &900 &HD283886\\
(347449)                & JFC-ACO & IRTF & 09/03/2016 &13:45 &1.1 &3.529 &2.724 &10.8  &15 &120 &HD142801\\
(366186)                & JFC-ACO & GTC & 17/12/2014 &02:16 &1.3 & 3.691 & 2.772 & 6.3 & 3 & 500 &SA 98-978\\
(380282)                & JFC-ACO & GTC & 17/12/2014 &06:09 &1.0 & 4.595 & 3.928 & 9.8 & 3 & 400 &SA 98-978\\
(397262)                & JFC-ACO & INT & 22/07/2015 &02:34 &1.1 &2.744  &1.797 &9.5  &3 &1200&HD206938\\
(405058)                & JFC-ACO & IRTF & 03/05/2016 &11:33 &1.3 &2.858 &1.869 &4.9 &36 &120 &SA 107-998\\
(465293)                & JFC-ACO & GTC & 17/12/2014 &01:49 &1.4 & 2.262 & 1.379 & 14.1 & 3 & 300 &SA 98-978\\
2002 RP$_{120}$ & Damocloid & INT & 17/09/2002 &05:31 &1.2 & 2.482 & 1.988 & 22.7 & 5 & 900 &Hyades 64\\
2002 RP$_{120}$ & Damocloid & TNG & 20/09/2002 &05:17 &1.3 & 2.480 & 1.928 & 22.1 & 6 & 240 &Hyades 64\\
2005 NA$_{56}$  & JFC-ACO & INT & 07/02/2015 &05:32 &1.4 &1.608 &0.627 &6.1   &4 &800 &HD79078\\
P/2006 HR$_{30}$        & Damocloid & WHT & 25/07/2006 &01:49 &1.1 & 2.413 & 1.618 & 18.3 & 3 & 900 &SA 112-1333\\
2006 EX$_{52}$  & Damocloid & WHT & 25/12/2006 &20:49 &1.1 & 2.626 & 1.789 & 13.6 & 2 & 600 &Hyades 64\\
2007 DA$_{61}$  & Damocloid & WHT & 10/03/2007 &00:02 &1.2 & 2.659 & 1.810 & 13.5 & 3 & 900 &SA 102-1081\\
2009 FW$_{23}$  & Damocloid & WHT & 10/04/2009 &21:16 &1.3 & 1.783 & 0.960 & 25.2 & 3 & 300 &SA 102-1081\\
\hline
\multicolumn{12}{l}{$^1$ Column \# is the number of exposures in the case of visible observations and the number of of individual}\\
\multicolumn{12}{l}{exposures taken in position A and B (AB) in the case of NIR observations.}\\
\multicolumn{12}{l}{$^2$ Integration time of individual exposures (or AB in the case of NIR observations).}
 \end{tabular}
\end{table*}

\subsection{Visible spectra of ACOs}

\subsubsection{10.4m GTC observations}
Low-resolution visible spectra of five ACOs were obtained using the Optical System for Imaging and Low Resolution Integrated Spectroscopy (OSIRIS) camera spectrograph \citep{cepa00,cepa10} at the 10.4m GTC.  OSIRIS has a mosaic of two Marconi 2048 x 4096 pixels CCD detectors,  a total unvignetted field of view of 7.8 x 7.8 arcmin, and a plate scale of 0.127 "/pix. To increase the signal-to-noise ratio (S/N) we selected the 2 x 2 binning and the standard operation mode with a readout speed of 200 kHz (with a gain of 0.95 e-/ADU and a readout noise of 4.5 e-). The tracking of the telescope was at the asteroid proper motion.

GTC spectra were obtained in service mode using the  R300R grism in combination with a second-order spectral filter that produces a spectrum in the 4800 to 9000 \AA~ with a dispersion of 7.74 \AA/pix for a 0.6" slit width. We used a 5.0" width slit oriented in parallactic angle to account for possible variable seeing conditions and minimize losses due to atmospheric dispersion. Series of three spectra were taken for all the targets. Consecutive spectra were shifted in the slit direction by 10" to correct for fringing. Observational details are listed in Table \ref{table1}.

\subsubsection{4.2m WHT observations}

Low-resolution visible spectroscopy of four ACOs were obtained using the double armed Intermediate dispersion Spectrograph and Imaging System (ISIS) at the 4.2m WHT. The R300B grating centered at 4.600\AA~  was used in the BLUE arm of ISIS  covering the 0.38 $-$ 0.54 $\mu$m region with a dispersion of 0.86 \AA/pixel for a 1" slit width. The R158R grating centered at 7.500\AA~ was used in the RED arm of ISIS  covering the 0.50 $-$ 0.95 $\mu$m region with a dispersion of 1.63 \AA/pixel. A 2" slit width was used oriented in the parallactic angle.  The tracking of the telescope was at the asteroid proper motion. Three spectra of each object were obtained by shifting the object 5" in the slit direction to better correct the fringing in the RED arm. 

\subsubsection{2.5m INT observations}
We obtained low-resolution visible spectroscopy of six ACOs using the Intermediate Dispersion Spectrograph (IDS) at the 2.5m INT.  The instrument was used with the $R150V$ grating and the $RED+2$ camera. This configuration allows one to cover the spectral interval 0.45-0.9 $\mu$m with a resolution of $R\sim550$ (for a 1" slit width). All observations were conducted by orienting the slit along the parallactic angle to minimize the effects of atmospheric differential refraction. The median ORM site seeing is 0.8" and the INT typical seeing is 1.2"; therefore we selected a 1.5" slit width.  Most asteroids were observed as close to the zenith as possible. The tracking followed the asteroid proper motion. 

\subsubsection{Data reduction and obtention of reflectance spectra}

Visible data reduction was done using standard image Reduction and Analysis Facility (IRAF) procedures. Images were over-scan and bias corrected, and flat-field corrected using lamp flats. The two-dimensional spectra were extracted and were collapsed to one dimension following sky background subtraction. The wavelength calibration was done using Ne and Ar lamps in the case of WHT and INT observations and Xe+Ne+HgAr lamps in the case of GTC observations.
Finally, the three spectra of the same object were averaged to obtain one final spectrum of the asteroid. 

\begin{figure}
\centering
  \includegraphics[width=9cm, angle=0]{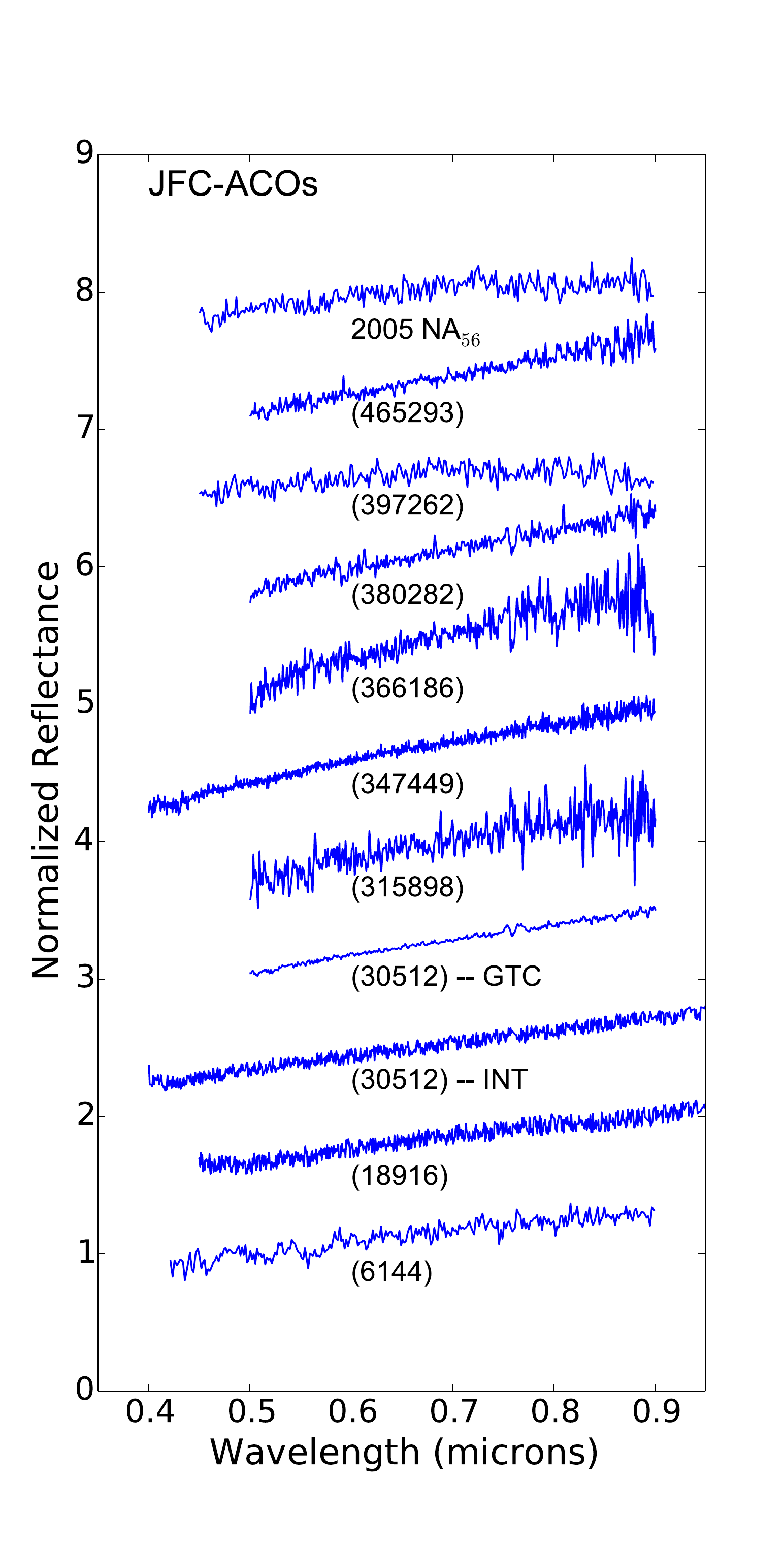}
\caption{Visible reflectance spectra of the JFC-ACOs presented in this paper, normalized to unity at 0.55 $\mu$m.  Spectra are shown with an offset in reflectance for clarity.}\label{visJFCACOs}
\end{figure}

\begin{figure}
\centering
  \includegraphics[width=9cm, angle=0]{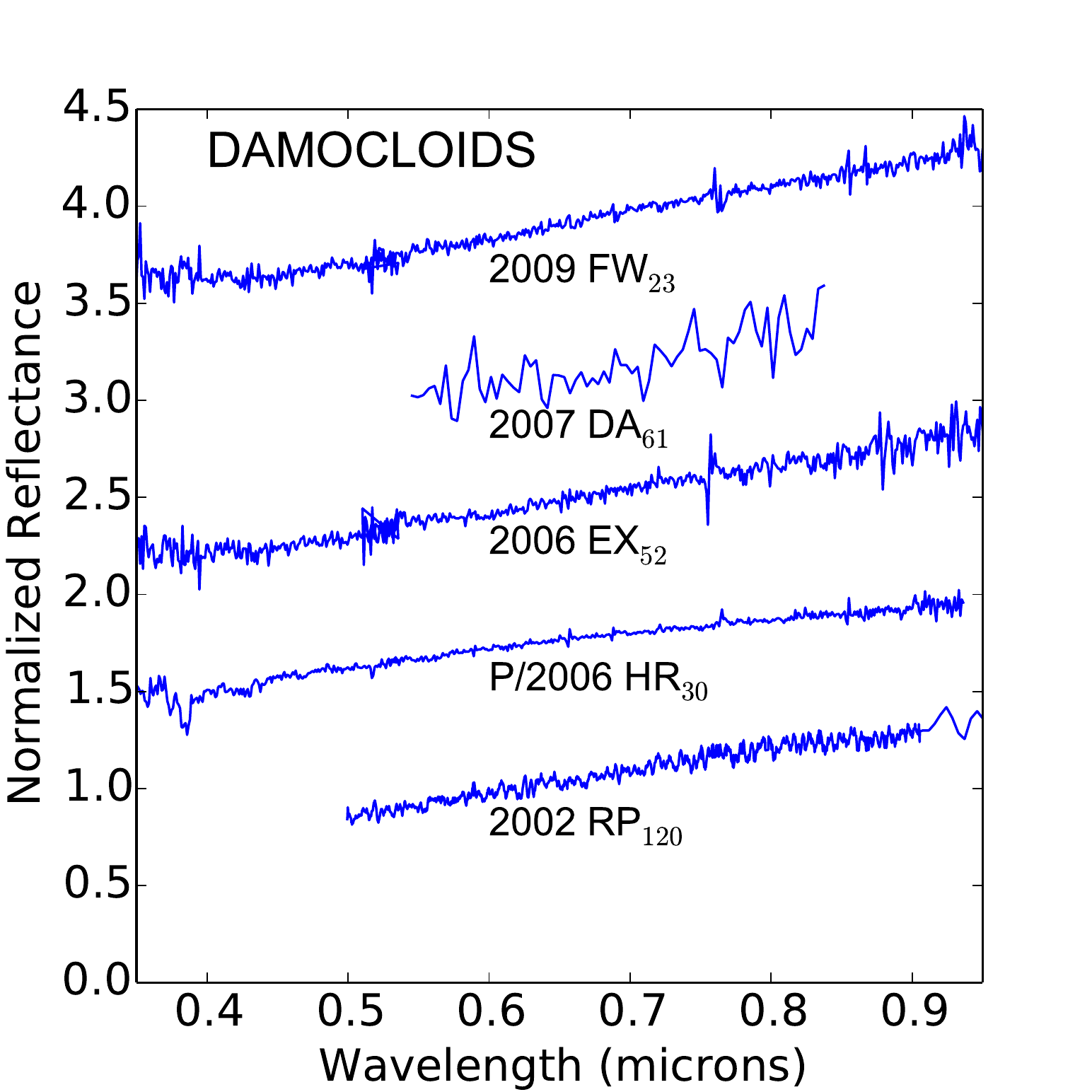}\caption{Visible reflectance spectra of the Damocloids presented in this paper, normalized to unity at 0.55 $\mu$m.  Spectra are shown with an offset in reflectance for clarity.}\label{visJFCDamo}
\end{figure}

At least one $G2V$ star from the Landolt catalog \citep{landolt92} was observed each night using the same spectral configuration and at similar airmass as that of the asteroids. These were used as solar analogs to correct for telluric absorptions and to obtain relative reflectance spectra.  The averaged spectrum of the ACO was divided by that of the solar analog, and the result was normalized to unity at 0.55 $\mu$m to obtain the reflectance spectrum. When possible, more than one $G2V$ star was observed in order to improve the quality of the final ACO reflectance spectra and to minimize potential variations in spectral slope introduced by the use of one single star. The solar analogs used to obtain the reflectance of each asteroid are shown in Table \ref{table1}.  When more than one solar analog was observed, we divided the spectrum of the asteroid by the spectra of the stars and checked against any possible variations in spectral slopes, which were of the order of 0.6\%/1000 \AA. We typically considered any variation smaller than 1\%/1000 \AA \  as good. Finally, an 11-pixel binning was applied to GTC spectra. The final reflectance spectra in the visible region of JFC-ACOs and Damocloids are shown in Figs. \ref{visJFCACOs} and  \ref{visJFCDamo}, respectively.

\subsection{Near-infrared spectra of ACOs}

\subsubsection{3.56m TNG observations}
We obtained low-resolution NIR spectroscopy of Damocloid 2002  RP$_{120}$ covering the 0.8-2.4 $\mu$m spectral range using the  3.56m TNG. We used the low-resolution mode of Near Infrared Camera Spectrograph (NICS), based on an Amici prism disperser that covers the 0.8-2.4 $\mu$m region  \citep{oliva01}. The slit was oriented in the parallactic angle and the tracking was at the asteroid proper motion. The width of the slit used was 1.5'' and corresponds to a spectral resolving power R $\approx$ 34 quasi-constant along the spectra. Observational details are listed in Table \ref{table1}.

The observational method and reduction procedure followed \citet{licandro2001}. The acquisition consisted of a series of short exposure images in one position of the slit (position {\em A}) and then offsetting the telescope by $10"$ in the direction of the slit (position {\em B}), and obtaining another series of images. This process was repeated and three series of {\em ABBA} cycles were acquired. The total on-object exposure time is 1440s. 

\subsubsection{3.0m IRTF observations}
Near-infrared spectral measurements of five ACOS in the 0.8-2.5 $\mu$m range were obtained using the SpeX instrument on the IRTF.  We used the low-resolution  Prism  mode  (R $\approx$ 100)  of  the  spectrograph \citep{2003PASP..115..362R}. Both ACOs and G2V stars were observed using the same configuration, with the 0.8" slit width aligned in parallactic angle. We followed the same observational method we used with the TNG. 
Several AB pairs of images were taken for each object, with individual image exposures typically being 120 s. Observational details are listed in Table \ref{table1}.

\subsubsection{Data reduction and obtention of reflectance spectra}

To correct for telluric absorption and to obtain the relative reflectance, a solar analog star or a $G2V$ star from the Landolt catalog \citep{landolt92} were observed during the same night at an airmass similar to that of the object (see Table \ref{table1}). 

\begin{figure}
\centering
  \includegraphics[width=9cm, angle=0]{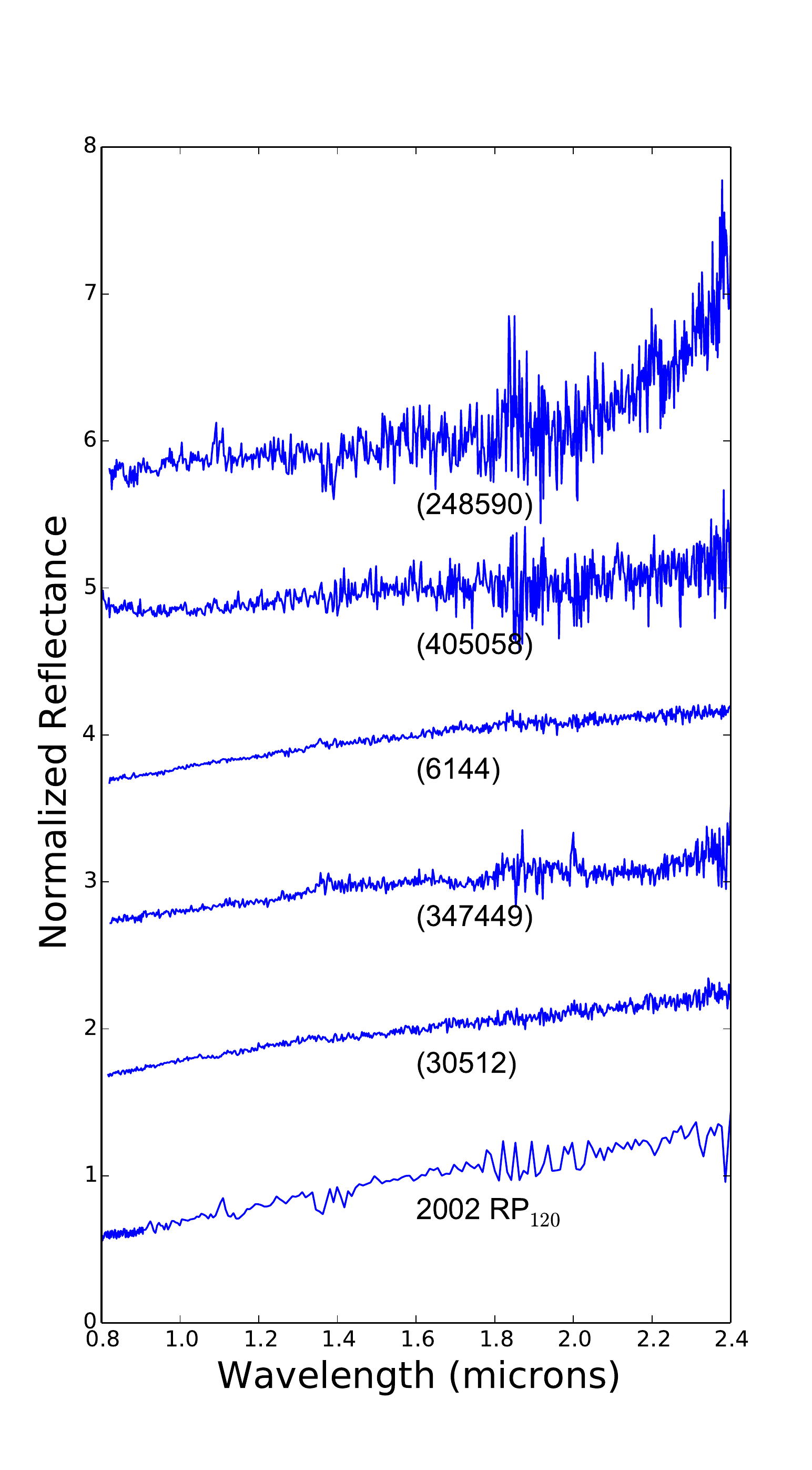}
\caption{NIR reflectance spectra of the ACOs presented in this paper, normalized to unity at 1.6 $\mu$m.  Spectra are shown with an offset in reflectance for clarity. }\label{irACOs}
\end{figure}

The two-dimensional spectra of ACOs and stars were extracted, and collapsed to one dimension. In the case of the TNG data, the wavelength calibration was performed using a look-up table which is based on the theoretical dispersion predicted by ray-tracing and adjusted to best fit the observed spectra of calibration sources and telluric absorptions. Finally, the spectra of the object were divided by the spectrum of the solar analog star, and the so obtained reflectance spectra averaged. Sub-pixel offsetting was applied when dividing the two spectra to correct for errors in the wavelength calibrations due to instrumental flexure. IRTF data reduction was performed similarly to what was done for TNG data using the new Spextool pipeline \citep{2004PASP..116..362C} for SpeX to create the final reflectance spectra of the ACOs. These spectra have been normalized to unity at 1.6 $\mu$m and are shown in Fig. \ref{irACOs}.

\section{Spectra of ACOs in the literature \label{sec:literature}}

To increase the sample for the statistical analysis of the ACO populations, we collected spectra of objects of the Tancredi's list published in the literature. From the  Small Main-Belt Asterids Spectroscopic Survey (SMASS, \citealt{bus2002tax}) we gathered visible spectra of five ACOs (two Damocloids and three JFC-ACOs). Near-infrared spectra of another three JFC-ACOs, and visible and NIR spectra of another two JFC-ACOs were gathered from MIT-UH-IRTF Joint Campaign for NEO Reconnaissance (http://smass.mit.edu/minus.html). The spectra of six of the ACOs retrieved from the MIT-UH-IRTF archive were analyzed by \citet{DeMeo2008}. From \citet{licandro08},  we also gathered visible and NIR spectra of two JFC-ACOs. The main orbital elements of all these objects are shown in Table \ref{table:elements2}.

\begin{figure}
\centering
  \includegraphics[width=9cm, angle=0]{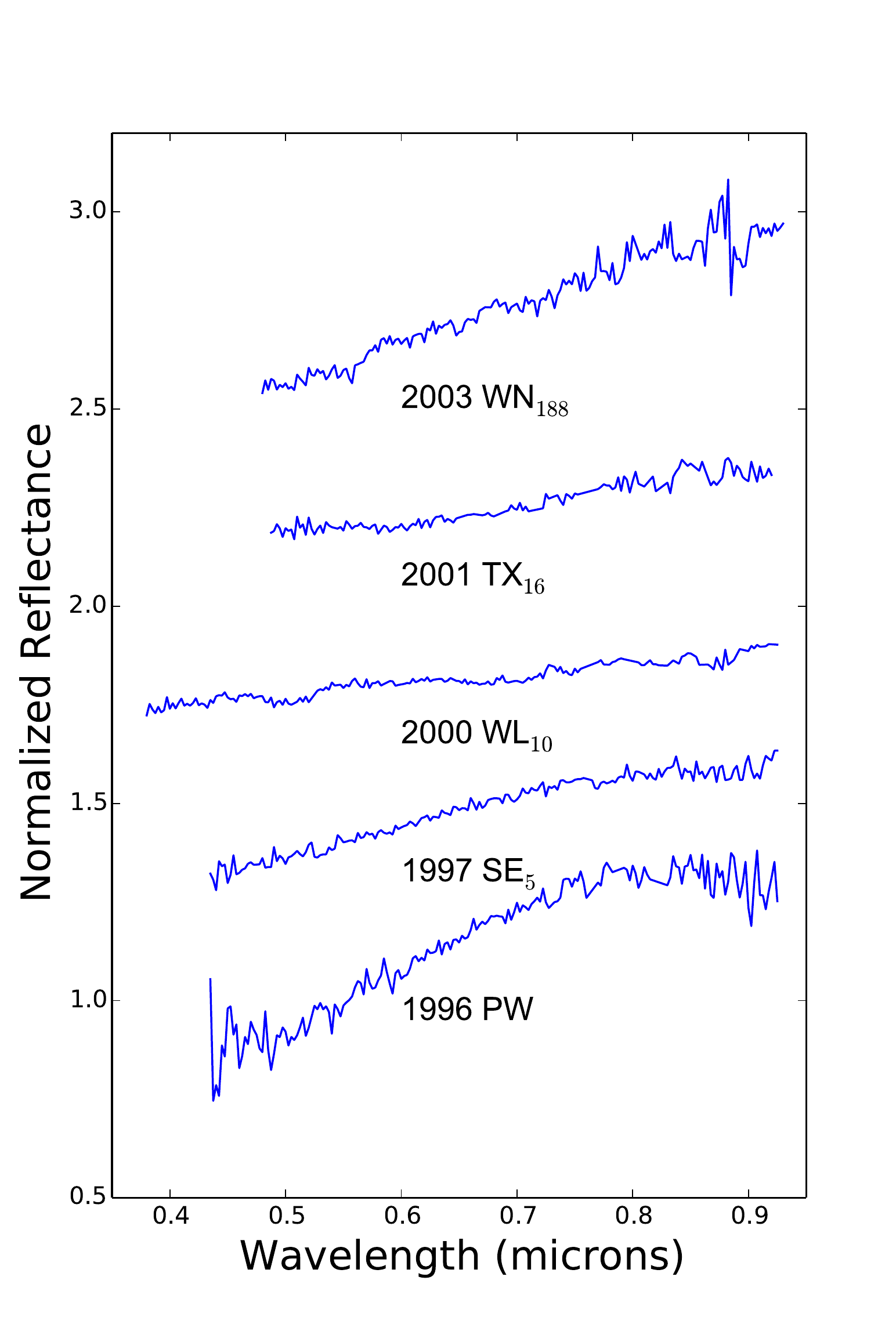}\caption{Visible reflectance spectra of the ACOs obtained from the literature, normalized to unity at 0.55 $\mu$m. Spectra are shown with an offset in reflectance for clarity. The two spectra in the lower part of the plot are the only classified as non-promitive X- or D-type; 1997 SE$_5$ corresponds to a T-type and 1986 PW to an L-type (see Sect. \ref{subsec:tax})}\label{visACOsLit}
\end{figure}

\begin{figure}
\centering
  \includegraphics[width=9cm, angle=0]{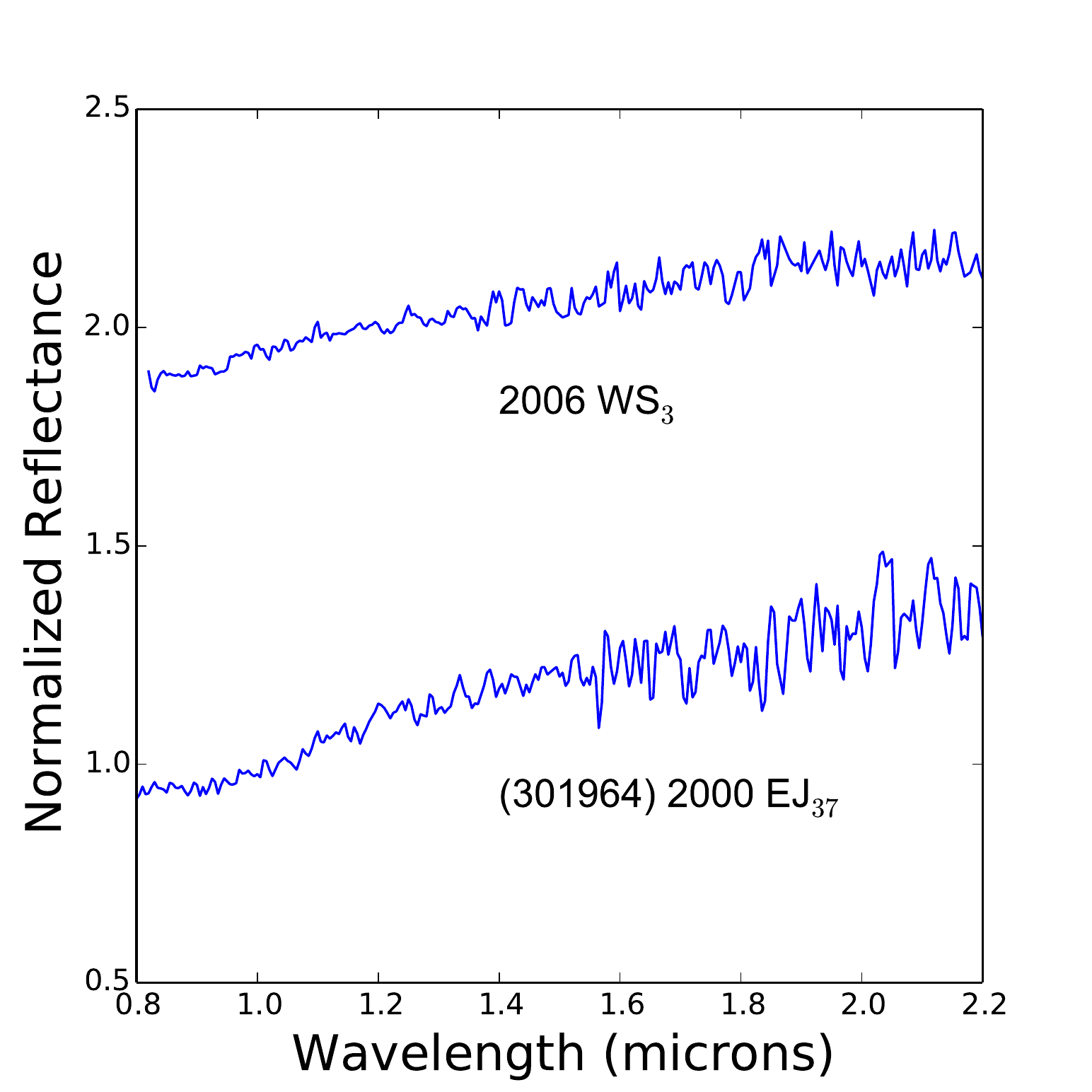}\caption{NIR reflectance spectra of the ACOs obtained from the literature, normalized to unity at 1.0 $\mu$m.  Spectra are shown with an offset in reflectance for clarity. }\label{irACOsLit}
\end{figure}

\begin{figure}
\centering
  \includegraphics[width=9cm, angle=0]{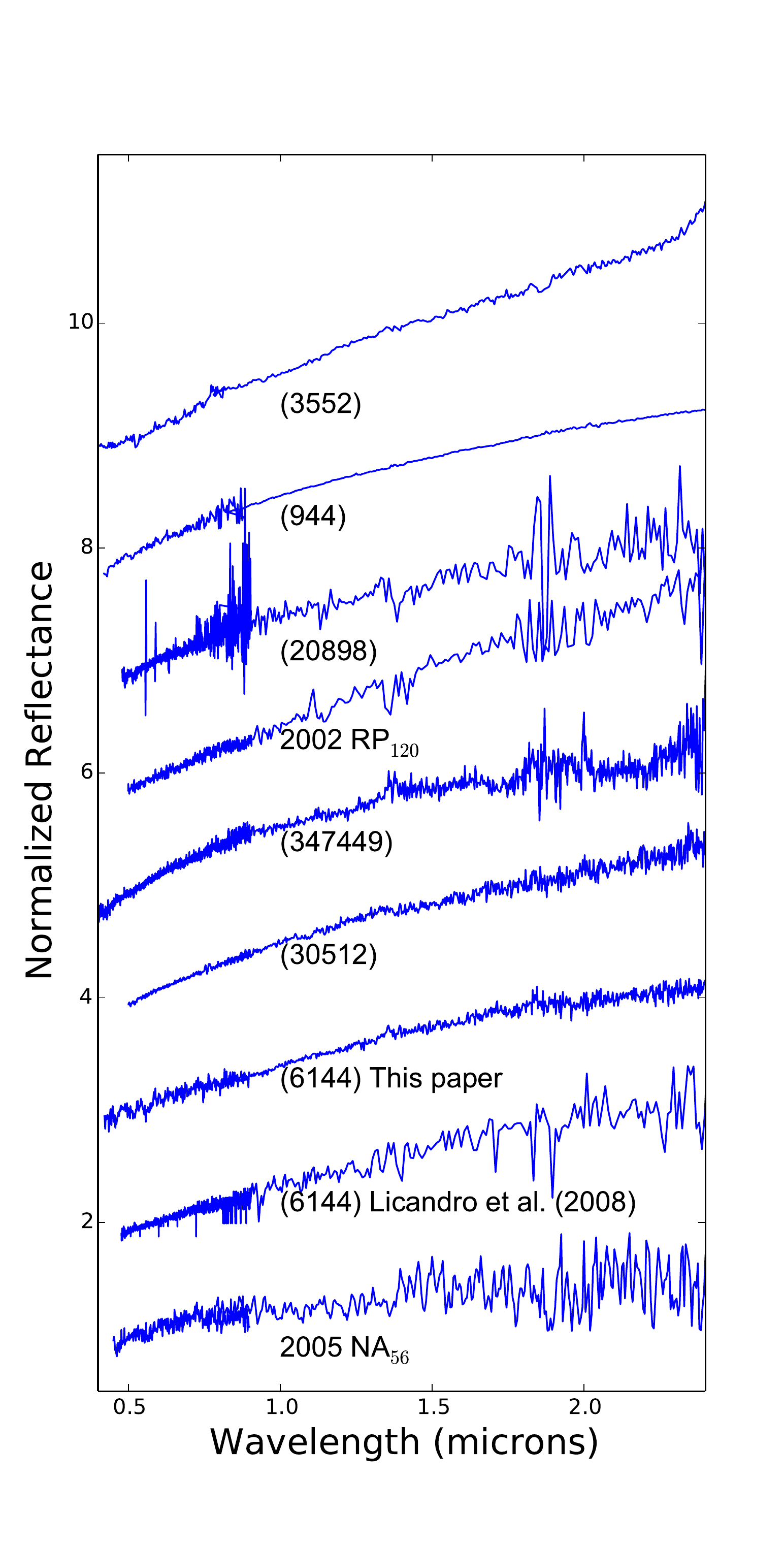}
\caption{Combined visible and NIR reflectance spectra of those ACOs for which data is available in both wavelength ranges. Spectra are normalized to unity at 0.55 $\mu$m and shown with an offset in reflectance for clarity. }\label{visnirACOs}
\end{figure}

Figure \ref{visACOsLit} shows the spectra of ACOs obtained from the literature with observations only in the visible range. Figure \ref{irACOsLit} shows the spectra of ACOs obtained from the literature with observations only in the NIR range. For those objects with both visible and NIR spectra, either obtained in this work or taken from the literature, we combined the two wavelength ranges using the common interval (0.8-0.9 $\mu$m), and the resulting complete spectra are shown in Fig. \ref{visnirACOs}.
                
\begin{table*}
\caption{\label{table:elements2} Orbital elements of the ACOs with spectra obtained from other sources taken from the NASA JPL Horizons ephemeris service (http://ssd.jpl.nasa.gov/horizons.cgi). {\it Group} is the dynamical subclass of the object.}           
\vskip4mm
\centering
\begin{tabular}{lcrcrcc}     
\hline
ACO & Group & $a$ (au) & $e$ & $i$ ($^{\circ}$)& $T_J$ & Source\\ \hline
(944) Hidalgo                   & JFC-ACO & 5.741 &0.661 & 42.52 & 2.068 &2 \\
(3552) Don Quixote              & JFC-ACO & 4.259 &0.709 & 31.09 &2.315  &2\\
(6144) Kondojiro                & JFC-ACO & 4.757 &0.362 & 5.88 & 2.867 &1 \\
(20898) Fountainhills           & JFC-ACO & 4.223  &0.465  &45.51  &2.349  &1 \\
1996 PW                         & Damocloid & 251.904 &0.990 &29.80  &1.724  &2  \\
1997 SE$_{5}$                   & JFC-ACO & 3.761 &0.566 & 92.55 & 2.655 &2 \\
301964 (2000 EJ$_{37}$ )                & JFC-ACO &4.631  &0.705 &10.06  &2.441  &2 \\
2000 WL$_{10}$          & JFC-ACO &3.143  &0.717 &0.25  &2.722   &2\\
2001 TX$_{16}   $               & JFC-ACO &3.579  &0.598 &8.14  &2.770   &2\\
2003 WN$_{188}$                 & Damocloid &14.384  &0.848 &26.96  &1.934   &2\\
2005 NA$_{56}   $               & JFC-ACO &3.029  &0.544 &9.14  &2.982  &2\\
2006 WS$_{3}    $               & JFC-ACO &3.468  &0.540 &29.99  &2.690   &2\\
\hline
\multicolumn{6}{l}{Source: (1) is Licandro et al. 2008; (2) is SMASS}\\
\end{tabular}
\end{table*}

\section{Results and analysis \label{sec:results}}

In this section we analyze the spectral properties of the ACO population using our own observations presented in Sect. \ref{sec:obs} and the spectra obtained from the literature shown in Sect. \ref{sec:literature}. We analyze the ACO's spectral properties as a whole, and then as two subpopulations: the JFC-ACOs and the Damocloids. We first analyze the taxonomical distribution of these objects following the \cite{demeo2009tax} scheme. Then, as almost all the spectra correspond to that of primitive asteroid classes, X- and D-types, for the characterization of the spectra we compute the visible and NIR spectral slope. Finally we look for the typical features due to the presence of hydrated minerals on the surface of asteroids: the 0.7 $\mu$m absorption band and the UV drop below $\sim$0.5 $\mu$m as we did in \citet{Depra2017}.

\subsection{Taxonomical distribution}\label{subsec:tax}

To perform the taxonomic classification, we used the on-line tool for modeling spectra of asteroids, M4AST \footnote{http://m4ast.imcce.fr/} \citep{popescu2012mast}. We first apply a polynomial fit to the asteroid spectrum, with varying order. Subsequently, the tool compares this fit at the corresponding wavelengths to templates of each taxonomical class defined by the \citet{demeo2009tax} taxonomy.  In cases where the spectrum is only in the visible range, the comparison is made to the templates defined in the \cite{bus2002tax} taxonomy. The adopted taxonomic class is the one with the smallest chi-squared. Results are shown in Table \ref{table:sparamJFCACOs} for the JFC-ACOs and in Table \ref{table:sparamDamocloids} for the Damocloids.  In the observed JFC-ACOs population, 11 objects (55\%) are D-types, 8 (40\%) are X-types and 1 (5\%) is T-type, while in the Damocloid population there are 6 (86\%)  D-types and 1 (14\%) L-type. 

Only two ACOs present spectra that do not correspond to a primitive X or D spectral class: one T-type (1997 SE$_5$) and one L-type (1996 PW); their spectra are shown in Fig. \ref{visACOsLit}. In the case of 1997 SE$_5$, the spectrum is very similar to that of a D-type with a very small change in slope above ~0.8 $\mu$m.  Considering also that the average albedo of T-type asteroids reported by \citet{Mainzer2011} is $p_V= 0.042,$ we can conclude that 1997 SE$_5$ is likely of primitive class.  In the case of 1996 PW, this change in spectral slope is larger, suggesting that it is more likely a non-primitive asteroid. On the other hand, there is published spectrophotometry by \citet{Hicks2000} using the Bessel B,V,R,I filters suggesting that it is compatible with a D-type asteroid with a reflectivity gradient slope S' =  11.3 $\pm$ 1.5 \%/1000\AA ~ and an unpublished visible spectrum by \citet{Hammergren1996}  that reported a "featureless red-sloped continuum with a reflectivity gradient S' = 10.2 $\pm$ 0.4 \%/1000\AA ~ over the wavelength range 620-860 nm", also consistent with the characteristics of a D-type asteroid. Unfortunately, there is no NIR spectra or albedo of these two objects that could help us to confirm whether these are rocky asteroids or dormant comets.

Another interesting case is that of (301964) 2000 EJ$_{37}$. Classified as X-type, its spectrum, shown in Fig.  \ref{irACOsLit}, presents a weak feature around 1$\mu$m similar to the band observed in some M-type asteroids \citep{Hardersen2005}. This band is attributed to low-Fe, low-Ca orthopyroxene minerals and has not been reported in the spectra of comet nuclei. M-type is one of the taxonomical classes in \cite{Tholen1984}  that is not defined in  \citet{demeo2009tax} taxonomy. M-types are included in the X-type classification because they present very similar spectra, and are distinguished from X-type because of their moderate geometric albedos ($0.07 < p_V < 0.30$). 2000 EJ$_{37}$ has a low albedo, $p_V = 0.049 \pm 0.009$ \citep{licandro16}, typical of that of a comet nucleus and too low for an M-type asteroid, confirming it as a primitive class object.

In the case of Damocloids, we can also consider the B,V,R,I spectrophotometric results of 12 objects in \citet{Jewitt2005}, three of them later observed active and classified as comets, and 11 Damocloids in  \citet{Jewitt2015b}. Both \citet{Jewitt2005} and \citet{Jewitt2015b} found that all the observed objects presented visible colors compatible with those of D-type asteroids.  

We conclude that, while almost 100\% of the Damocloids are D-types, JFC-ACOs are $\sim$ 60\% D-types and $\sim$ 40\% X-types.

\subsection{Distribution of spectral slopes ($S'$)}\label{slopes}

We compute the spectral slope $S'$ in the 0.55-0.86 $\mu$m wavelength range following the $S'$ definition in \citet{LuuJew1996}. In the large majority of cases, the reflectance is well represented by a linear fit. The fit is normalized to unity at 0.55 $\mu$m and $S'$ is computed in units of  \%/1000\AA . To compute  the NIR slope, we use the reflectance in the 1.0-1.75 $\mu$m spectral range,  with the spectrum normalized to unity at 1.2 $\mu$m. In this way, we use the region with better S/N and avoid any possible thermal emission contamination like that seen in the spectrum of JFC-ACO (248590) 2006 CS (see Fig. \ref{irACOs}). The spectral slopes computed for JFC-ACOs and Damocloids are shown in Tables \ref{table:sparamJFCACOs} and \ref{table:sparamDamocloids}, respectively. The $S'$ uncertainty is largely dominated by systematic errors rather than the errors of the linear fit. These errors can be evaluated when several solar analog stars are observed during the observing night and are usually in the order of 1 -- 0.5 \%/1000\AA ~ in the visible and $\sim$ 1 \%/1000\AA ~ in the NIR. We assume these values as the uncertainties of the computed slopes instead of using the error of the linear fit.

\begin{figure}
\centering
  \includegraphics[width=9cm, angle=0]{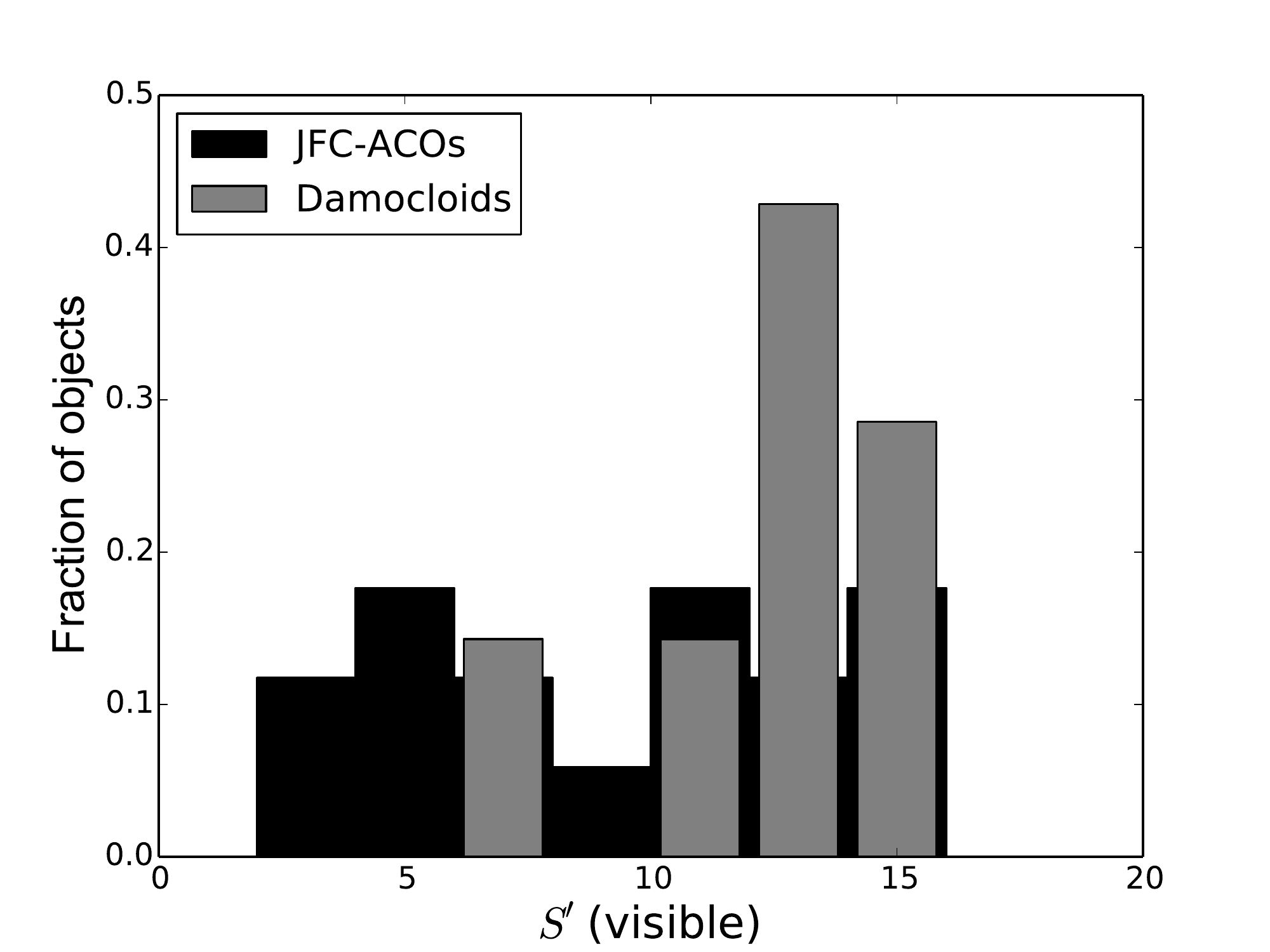}\\
\caption{Distribution of spectral slopes of JFC-ACOs and Damocloids in the visible spectral range.}\label{Fig:histS}
\end{figure}
                
\   In the visible, the mean $S'$  of JFC-ACOs is 9.7 $\pm$ 4.6 \%/1000\AA\  and for the Damocloids this is 12.2 $\pm$ 2.4  \%/1000\AA\  (see Fig. \ref{Fig:histS});  in the NIR,  $S'$  is  3.3 $\pm$ 1.0 \%/1000\AA ~for the JFC-ACOs, but we have only one Damocloid with available NIR spectrum. The differences in $S'$ computed in the visible and NIR are explained in \citet{licandro08} and are due to the different wavelength used to normalize the visible and NIR spectra and the typical curvature of D- and X-type objects (see Fig. \ref{visnirACOs}). The $S'$ distribution of Damocloids that we obtained is very similar to that in \citet{Jewitt2005} derived using $B,V,R,I$ color photometry of 12 Damocloids, with a mean $S'$ = 11.9 $\pm$ 1.0  \%/1000\AA.  \citet{Jewitt2005} obtained $B,V,R,I$ color photometry of another 11 Damocloids and derived a slightly redder mean color that was transformed to $S'$ using Eq. 2 in \citet{Jewitt2002} resulting in a mean $S' = 13.3 \pm$ 2.0  \%/1000\AA, also compatible with the slope distribution derived in this paper.

The differences in the slope distribution between JFC-ACOs and Damocloids is what we expect from the different distribution of spectral classes shown in the previous section. Is this an observational artifact? The only known observational effect that can change the color of an asteroid is the phase reddening \citep{Gradie1980,Gradie1986} where the spectral slope $S'$  becomes redder with increasing phase angle ($\alpha$). Looking at Table \ref{table1}, we notice that all but one of our observations were done at $\alpha <26^{\circ}$, so no large phase reddening effects are expected. In fact, the object observed at $\alpha = 89^{\circ}$, one of the least red objects in our sample, is classified as X-type. We conclude that the differences in spectral slope distribution between both populations are real and should be explained as differences in the composition and/or space weathering effects.

\begin{figure}
\centering
  \includegraphics[width=9cm, angle=0]{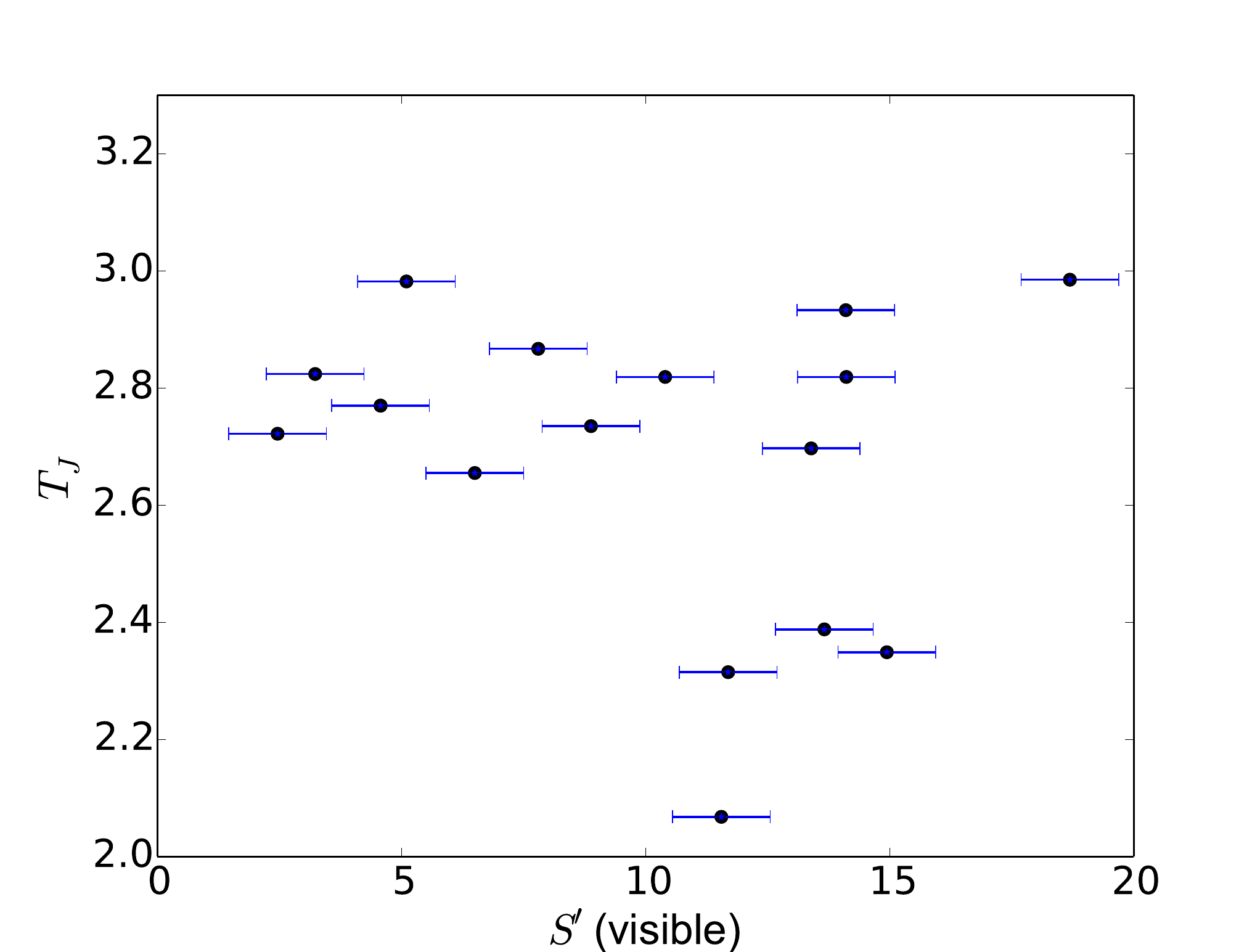}\\
\caption{$S'$ vs. Tisserand parameter of the JFC-ACOs population.}\label{TvsS}
\end{figure}

On the other hand, \citet{licandro08} reported a correlation between $T_J$ and $S'$ in the JFC-ACOs population (see their Fig. 10), the smaller the $T_J$, the larger the $S'$.  In the sample of JFC-ACOs presented in this work, this relation is not present anymore (see Fig. \ref{TvsS}); there are several objects with $2.6 < T_J < 3.0$ with rather high values of $S'$. We still did not find any object with $T_J < 2.6$ and $S' < 10$ \%/1000\AA; although the sample is  not large enough (only 4 objects with $T_J < 2.6$) to make a firm conclusion. Comparing also our Fig. 8 with Fig. 10 in \citet{licandro08}, we conclude that Tancredi's criteria discard many objects with $T_J \sim 3.0$  and $S' < 3$ (likely C-, B- type interloper asteroids, not dormant comets).

Within the limits of the sample of ACOs presented here, we could not find any correlation of spectral slope with the orbital parameters or with the absolute magnitude ($H$).

\begin{table}[!hb]
\caption{Spectral slopes and taxonomical classification for the JFC-ACOs. Visible and NIR slopes were computed normalizing the spectrum at 0.55 and 1.20 $\mu$m, respectively. }
\label{table:sparamJFCACOs}
        \centering
        \footnotesize
        \begin{tabular}{ccclc} \hline 
        ACO &   Visible  & NIR  & Tax & Tel. \\ 
        . & slope       & slope &  & \\
         & (\%1000/\AA)  & (\%1000/\AA)  &  & \\
        \hline 
        (944)                   & 11.55         & 4.00          & D*            & 5       \\
        (3552)                  & 11.69         & 5.37          & D*            & 5       \\
        (6144)                  & 7.91  & -             & D             & 2       \\
        (6144)                  & 7.68  & 5.05          & D*            & 3       \\
        (6144)                  & -             & 4.40          & D             & 6       \\
        (18916)                         & 8.88  & -             & D             & 2       \\
        (20898)                         & 14.94         & 1.89          & D*              & 3     \\
        (30512)                         & 11.33         & -             & D               & 1     \\
        (30512)                         & 9.49  & -             & D             & 2       \\
        (30512)                         & -             & 3.98          & D               & 6 \\
        (248590)                        & -             & 2.22  & X             & 6       \\
        (301964)                        & -             & 3.36  & X             & 5       \\
        (315898)                        & 13.66         & -             & D               & 1     \\
        (347449)                        & 13.39         & -             & D               & 2     \\
        (347449)                        & -             & 3.46  & D             & 6       \\
        (366186)                        & 18.69         & -             & D               & 1     \\
        (380282)                        & 14.11         & -             & D               & 1     \\
        (397262)                        & 3.23  & -             & Xk    & 2       \\
        (405058)                        & -             & 2.80  & X             & 6 \\
        (405058)                        & 4.57  & -             & X             & 5       \\
        (465293)                        & 14.10         & -             & D               & 1     \\
        1997 SE$_5$             & 6.50  & -             & T             & 5       \\
        2000 WL$_{10}$  & 2.46  & -             & X             & 5     \\
        2001 TX$_{16}$  & 4.57  & -             & X             & 5     \\
        2005 NA$_{56}$  & 5.10  & -             & X*            & 2     \\
        2005 NA$_{56}$  & -             & 2.53  & X*            & 5     \\
        2006 WS$_3$             & -             & 2.23  & X             & 5       \\
        \hline
        \multicolumn{5}{l}{\footnotesize Telescope: (1) GTC, (2) INT (3) Licandro et al. (2008),}\\
        \multicolumn{5}{l}{\footnotesize         (4) WHT, (5) SMASS, (6) IRTF.  } \\
        \multicolumn{5}{l}{\footnotesize (*) Taxomony classification made with combined}\\
        \multicolumn{5}{l}{\footnotesize         visible and NIR data. }
        \end{tabular}
        \end{table}

\begin{table}[!hb]
\caption{Spectral slopes and taxonomical classification for the Damocloids. Visible and NIR slopes were computed normalizing the spectrum at 0.55 and 1.20 $\mu$m, respectively. }
\label{table:sparamDamocloids}
        \centering
        \footnotesize
        \begin{tabular}{ccclc} \hline 
        ACO &   Visible  & NIR  & Tax & Tel. \\ 
         & slope        & slope &  & \\
         & (\%1000/\AA)  & (\%1000/\AA)  &  & \\
        \hline \hline
        1996 PW                         & 12.53         & -             & L & 5   \\
        2002 RP$_{120}$         & 12.42         & 6.38          & D* & 3        \\
        2003 WN$_{188}$         & 10.24 & -             & D & 5 \\
        2006 EX$_{52}$  & 12.32         & -             & D & 4 \\
        P/2006 HR$_{30}$        & 7.96  & -             & D & 4 \\
        2007 DA$_{61}$  & 15.83         & -             & D & 4 \\
        2009 FW$_{23}$  & 14.23 & -             & D & 4 \\
        \hline
        \multicolumn{5}{l}{\footnotesize Telescope: (1) GTC, (2) INT (3) Licandro et al. (2008),}\\
        \multicolumn{5}{l}{\footnotesize         (4) WHT, (5) SMASS, (6) IRTF  } \\
        \multicolumn{5}{l}{\footnotesize (*) Taxomony classification made with combined}\\
        \multicolumn{5}{l}{\footnotesize         visible and NIR data. }
        \end{tabular}
        \end{table}


\subsection{Searching for hydration signatures}

The presence of aqueously altered minerals on asteroid surfaces can be inferred by a  shallow spectral absorption band centered at 0.7 $\mu$m. We search for the presence of this feature in our sample applying the same method presented in \citet{Depra2017}. First we calculate the continuum with a linear fit within the 0.55--0.58 and 0.83--0.86 $\mu$m intervals and divide the spectra by this continuum. Then we fit a fourth-order
spline in the 0.58--0.83 $\mu$m range and compute the depth and central wavelength of the absorption, if any. For the error estimation, we ran a Monte-Carlo model with 1000 iterations, randomly removing 20 \% of the points, and measuring the band depth at each iteration. The final value for the band depth is the center of the resultant distribution and the error is the variance.

Another indicator of hydration is a decrease in reflectance shortward of 0.5 $\mu$m typically observed in the spectra of primitive asteroids. This is associated with the presence of an ultra-violet (UV) absorption feature produced by iron-bearing silicates \citep{Vilas1995} on the surface of the asteroid. The decrease in reflectance is normally quite dramatic, implying a substantial change in spectral slope. The exact wavelength at which this drop in reflectance is occurring is called the  ``turn-off point''. To characterize the turn-off point, we first perform a linear fit using just ten points at the beginning and at the end of the spectra and then measure the distance from the data points in the spectra from the fit.  To investigate a possible turn-off in the spectrum we assume the condition that the distance of the farthest away point with respect to the corresponding value of the linear fit should be larger than 3.5\% . 

\begin{figure}
\centering
  \includegraphics[width=9cm, angle=0]{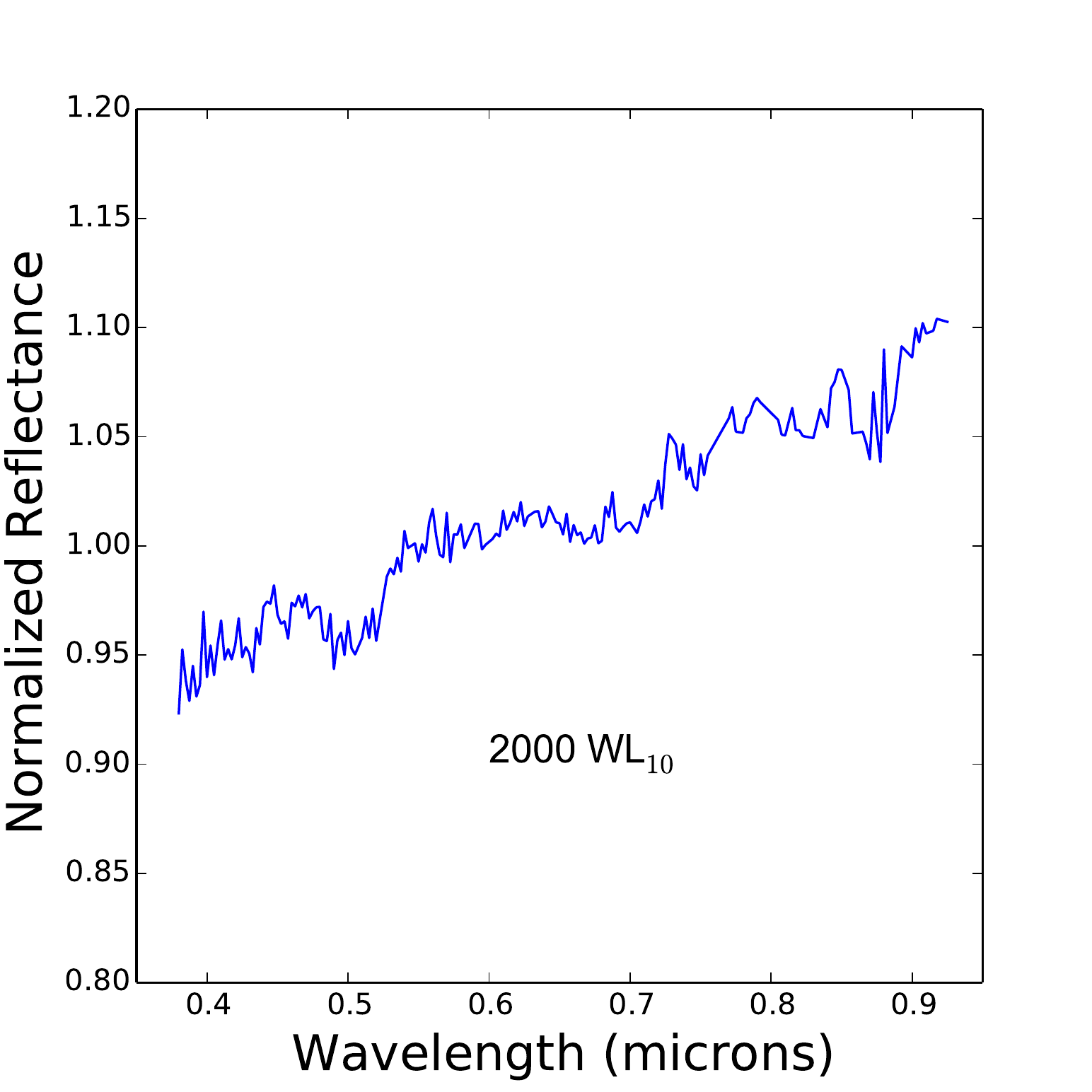}
\caption{Visible reflectance spectra of 2000 $WL_{10}$ obtained from the SMASS, normalized to unity at 0.55 $\mu$m.}\label{visWL10}
\end{figure}

Only one object presents the  $0.7 \mu$m band, 2000 WL$_{10}$, with a band depth of only 1.47 $\pm$ 0.14\% centered at 0.686 $\pm$ 0.004 $\mu$m. A further look at its published spectrum (see Fig. \ref{visWL10}) shows that it has a similar structure around $0.5 \mu$m and perhaps another around 0.9 $\mu$m, indicative of some problems with the spectrum itself instead of real absorption features.  
On the other hand, none of the observed ACOs present a measurable UV drop in the reflectance spectrum, although only ten of them were observed well below  $0.5 \mu$m. We conclude that there is no evidence of water hydration on the surface of ACOs from their visible spectra. Nonetheless, 
spectra with high S/N in the $3  \mu$m are needed to make a definitive statement about this. 

\subsection{Analysis of thermal excess in the spectrum of ACO (248590)}\label{sec:thermal}

The reflectance spectrum of (248590) 2006 CS (top of Fig.~\ref{irACOs}) shows a thermal excess due to the high surface temperatures prevailing at the time when this object was observed at a heliocentric distance of 0.88 au (Table~\ref{table1}). We examined this excess using the near-Earth asteroid thermal model (NEATM, \citealt{Harris1998}) implemented in \citet{AliLagoa2017}  and a reflected light model based on the $H$-$G$ values tabulated in the Minor Planet Center (e.g., see description in \citealt{AliLagoa2013}) modified to match the spectral slope measured in this work. 

\begin{figure}
\centering
  \includegraphics[width=9cm]{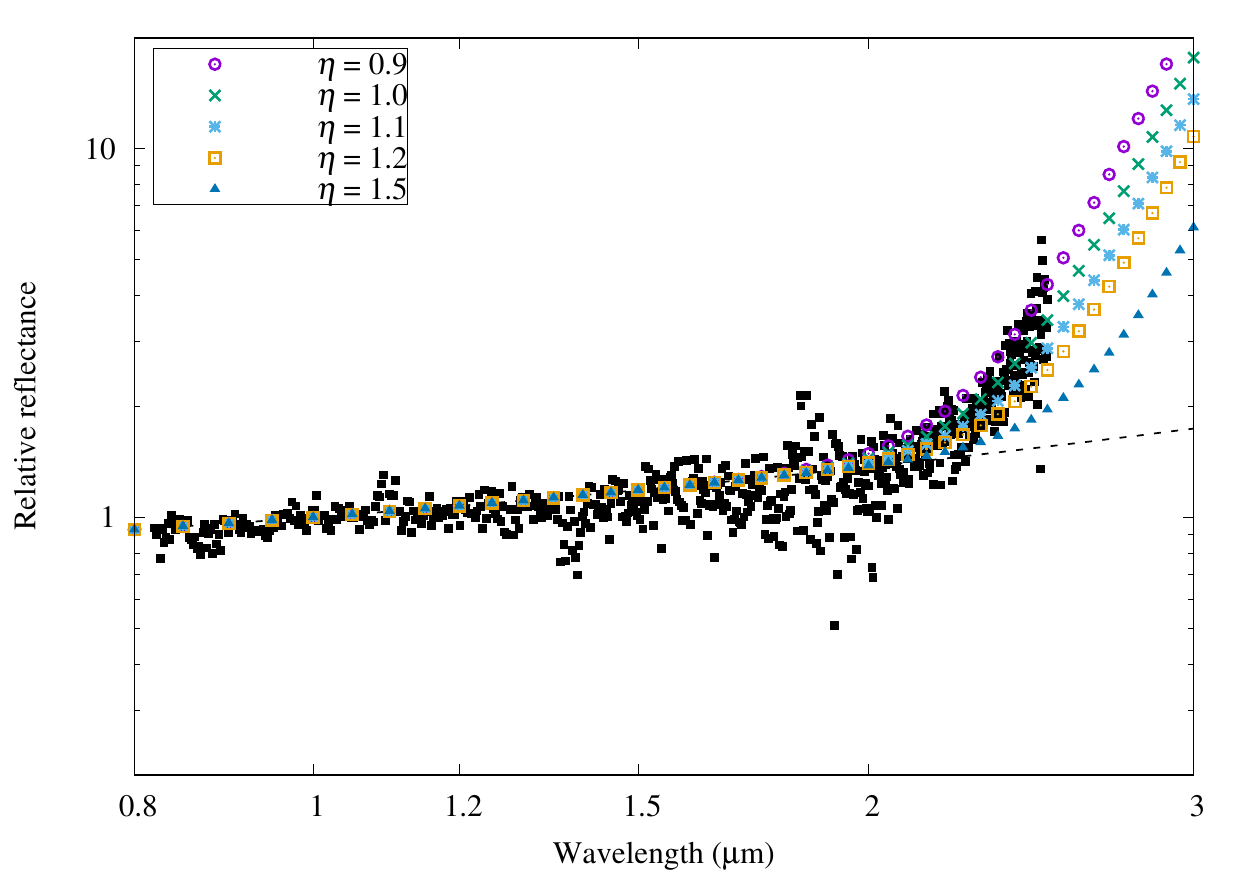}
\caption{Relative reflectance (black squares) and thermal and reflected light model (colored symbols) of asteroid (248590) 2006 CS normalized to unity at 1.0~$\mu$m. The dashed line shows the reflected light model without the thermal component. Values of the infrared beaming parameter ($\eta$) higher than 1.2 do not fit well the observed thermal excess.}\label{fig:thermal}
\end{figure}

Figure~\ref{fig:thermal} shows the observed reflectance spectrum and the models normalized to unity at 1~$\mu$m. We assumed the average value of the visible geometric albedo ($p_V$) of ACOs found by \citet{licandro16} and different beaming parameters ($\eta$), from 0.9 to 1.5. We find that only $\eta$-values in the range 0.9 to 1.2 reasonably match the observed thermal excess, which is puzzling since most objects observed at such high phase angles require $\eta$-values in the range 1.5 to 2.0 (e.g., see Mainzer et al. 2011 and the discussion in \citealt{AliLagoa2017} ). This means that we require higher-than-expected temperatures to correct for this thermal excess. A plausible explanation is that this object's thermal inertia is high enough for the non-illuminated part to contribute significantly to the observed flux on the evening side, which the NEATM compensates by increasing the temperature (lowering the $\eta$). However, this interpretation cannot be firmly established until we further characterize the rotational state and shape of this object. 

\section{Comparison with spectral properties of comet nuclei \label{sec:comparison}} 

A comparison between the spectral properties of ACOs and cometary nuclei is needed to study their possible cometary nature. If ACOs are dormant or extinct comets, it is expected that their surface is covered by a widespread insulating thick mantle of fine dust produced by past cometary activity. There should also be some very low active and almost inactive comets. In fact, between the observed ACOs,  low cometary activity was reported on 2006 HR$_{30}$  \citep{Hicks2007},  meaning that it is actually classified as comet P/2006 HR$_{30}$ (Siding Spring),  and a faint coma and tail where detected in the 4.5 $\mu$m Spitzer Space Telescope band images of (3553) Don Quixote  \citep{Mommert2014}.

Only a few visible or NIR  spectra of comet nuclei are published: 124P/Mrkos, 19P/Borrelly, C/2001 OG$_{108}$, 162P/Siding Spring and 28P/Neujmin (\citealt{Licandro2003, Soderblom2004,Abell2005,Campins2006,Campins2007,Filacchione2016}, respectively). All of them are featureless with a red slope in the 0.5 to 2.5 $\mu$m region, typical of X- or D-type asteroids and similar to the spectra of the ACOs presented in this paper (see e.g., Fig. 3 in \citealt{Campins2007} and Fig. 6 in \citealt{Filacchione2016}). No water-ice absorption bands are observed in the NIR spectra of comet nuclei, indicative that, at most, very little (a few percent) water ice is on their surface. \citet{Kelley2017} showed that the 5-35 $\mu$m spectra of these comets and comet 9P/Temple 1 present a weak 10-$\mu$m plateau, similar to the plateau observed in the thermal spectra of Jovian Trojan D-types (624) Hektor, (911) Agamemnon and (1172) Aneas, indicative of the presence of a dust layer. Actually, \citet{Emery2006} attributed the 10-$\mu$m plateau observed in D-type Trojans to a layer of fine dust in a fairy-castle structure. \citet{Licandro2011b,Licandro2012} and \citet{Vernazza2012}  showed that X- and C-type asteroids can also present an emission plateau in the 10 $\mu$m region,  usually less prominent than those observed in the D-types but indicative of a fine dust mantle covering their surface.  

There are also a few spectra of the so called Main Belt Comets (MBCs, \citealt{Hsieh2006})  published (e.g.,   \citealt{Licandro2011,Licandro2013}) and spectra of active asteroids (3200) Phaethon \citep{Licandro2007, deLeon2010} and 107P/Wilson-Harrington \citep{Ishiguro2011}. There is also a spectrum of the near-Earth Jupiter family comet 249P/LINEAR \citep{Fernandez2017}. All these objects present featureless and almost neutral spectra like those of asteroids of the C-complex. However, active asteroids (including MBCs) and 249P are likely formed in the asteroid main belt and not scattered from the Oort Cloud or the trans-Neptunian belt as is the population of ACOs studied in this work. Therefore, we do not consider them for this analysis. 

The color of comets and related populations has recently been analyzed in   \citet{Jewitt2015b}  using  spectrophotometric observations obtained from different sources including their own. In particular, Jewitt used the colors of cometary nuclei (16 JFCs and 5 LPCs) reported by \citet{Lamy2009} using the $B,V,R,I$ filters of the Hubble Space Telescope Planetary Camera, the colors of 31 active JFCs in \citet{Solontoi2012} based on Sloan Digital Sky Survey (SDSS) observations using the $u, g, r, i, z$ filters, and their own observations of 26 active LPCs using the Keck telescope and $B,V,R,I$ filters. \citet{Jewitt2015b} found that the colors of the dust in the coma and the comet nuclei are identical within the uncertainties and that JFCs and LPCs are significantly redder than the Sun. \citet{Lamy2009} obtained the mean colors of the nuclei of JFCs using their own HST observations and photometry obtained from the literature.  Table \ref{table:colorcomets} summarizes the $(B-V)$ and $(V-R)$ colors of the active comets and comet nuclei obtained from these three sources. To obtain the mean value and standard deviation of comet nuclei in \citet{Lamy2009} we used the adopted colors from their Table 5. To derive the colors of active comets in JFCs we used their $(u-g)$ and $(g-r)$ values transformed to $(B-V)$ and $(V-R)$ from equations derived for stars\footnote{http://www.sdss3.org/dr8/algorithms/sdssUBVRITransform.php}. The colors show that all the considered population present very similar mean colors (considering the standard deviations) while JFCs present a wider distribution than LPCs, similar to what we found between JFC-ACOs and Damocloids in Sect. \ref{slopes}. Photometry does not show the differences in mean spectral slope between the JFCs and LPCs that we observe with spectroscopy between JFC-ACOs and Damocloids, likely because the spectral slopes based on photometric measurements from different sources have larger uncertainties than spectroscopic ones, and the differences in the mean spectral slopes observed here are small (only 2.5\%).                                                                                                                                                                           

\begin{table}[!hb]
\caption{Colors of comets and ACO populations. The colors of comets are computed based on the data in column {\em Data source}. The colors of JFC-ACOs and Damocloids are computed from the visible $S'$ distributions obtained in this paper using Eq. 2 in \citet{Jewitt2002}.}
\label{table:colorcomets}
        \centering
        \footnotesize
        \begin{tabular}{lccr} \hline 
        Population &    $(B-V)$ & $(V-R)$ & Data source \\ 
        \hline \hline
        Active JFCs & 0.79 $\pm$ 0.11 & 0.44 $\pm$ 0.03 &       \citet{Solontoi2012}\\  JFCs nuclei & 0.82 $\pm$ 0.11 & 0.48 $\pm$ 0.10  &    \citet{Lamy2009}\\
        JFC-ACOs  & 0.76 $\pm$ 0.05 & 0.42 $\pm$ 0.06 & this paper      \\
        Active LPCs & 0.78 $\pm$ 0.02 & 0.47 $\pm$ 0.02 &        \citet{Jewitt2015b}\\  LPCs nuclei & 0.80 $\pm$ 0.07 & 0.43 $\pm$ 0.04 &     \citet{Lamy2009}\\
        Damocloids & 0.79 $\pm$ 0.03 & 0.47 $\pm$ 0.02 &        this paper\\
        Sun & 0.64 $\pm$ 0.02 & 0.35 $\pm$ 0.01 &        \\
        \hline
        \end{tabular}
        \end{table}

\section{Conclusions \label{sec:conclusions}} 

Visible and/or NIR spectra of the 17 ACOs selected using \citet{Tancredi2014} dynamical criteria were obtained using four different telescopes located at  "El Roque de los Muchachos Observatory" (Canary Islands, Spain), the 10.4m GTC, the 4.2m WHT, the 3.56m TNG and the 2.5m INT, and the 3m NASA Infrared Telescope Facility (IRTF, Mauna Kea, USA).  We also collected previously published spectra of 12 ACOs in the Tancredi's list.
All but two ACOs have a primitive-like class spectrum (X or D-type), one T-type (1997 SE$_5$) and one L-type (1996 PW).   Almost 100\% of the Damocloids are D-type, whereas JFC-ACOs are $\sim$ 60\% D-types and $\sim$ 40\% X-types. The mean spectral slope $S'$ of JFC-ACOs is 9.7 $\pm$ 4.6 \%/1000 \AA \ and for the Damocloids is 12.2 $\pm$ 2.4  \%/1000\AA  ~ in the visible,  and 3.3 $\pm$ 1.0 \%/1000\AA ~for the JFC-ACOs in the NIR. The differences in the slope distribution between JFC-ACOs and Damocloids is what we expect from the different distribution of spectral classes shown above and are not due to an observational phase reddening effect. The spectral slope and spectral class distribution of ACOs is similar to that of comets, which supports a cometary origin.

No correlations between spectral slopes and the orbital parameters or absolute magnitude ($H$) are found. The correlation between $T_J$ and $S'$ in the JFC-ACOs population reported by \citet{licandro08} is not observed in the sample of JFC-ACOs presented in this work (see Fig. \ref{TvsS}). There are several objects with $2.6 < T_J < 3.0$ with rather high values of $S'$ and there is no object with $T_J < 2.6$ and $S' < 10$ \%/1000\AA, but the sample of objects with $T_J < 2.6$ is still too small to be conclusive. Tancredi's criteria discard many objects with $T_J \sim 3.0$  with $S' < 3$ (likely C- or B-type interloper asteroids, no dormant comets) considered in \citet{licandro08}, supporting the idea that Tancredi's dynamical criteria for selecting dormant comet candidates are much more rigorous than those used before.

Finally, there is no evidence of water hydration on the surface of ACOs from their visible spectra, which is also compatible with a cometary origin.   Spectra of cometary nuclei  short-ward of 0.5 $\mu$m do not show an absorption toward the ultraviolet \citep{Abell2005,Campins2006} like those produced by hydrated minerals. Also, the dust of 67P/Churyumov-Gerasimenko is devoid of hydrated minerals \citep{Bardyn2017}.  

We conclude that the ACOs in \citet{Tancredi2014} are likely dormant comets considering dynamical and surface-properties criteria, their albedo distribution \citep{licandro16}, and spectral properties.


\begin{acknowledgements}
      We thank ING students F.~Lopez-Martinez, R.~Errmann, I.~Ordonez-Etxeberria and H.~F.~Stevance that conducted INT observations. We also thanks Dr. Driss Takir for his useful comments that helped to improve the manuscript. 
      J. Licandro, D. Morate, M. Popescu, and J. de Le\'on acknowledge support from the AYA2015-67772-R  (MINECO, Spain). JdL acknowledges support from from MINECO under the 2015 Severo Ochoa Program SEV-2015-0548
      The work of M. Popescu was also supported by a grant of the Romanian National Authority for  Scientific  Research --  UEFISCDI,  project number PN-II-RU-TE-2014-4-2199.    M. De Pr\'a acknowledges funding from the CNPq (Brazil).
     Based on observations made with the Gran Telescopio Canarias (GTC), the Italian Telescopio Nazionale Galileo (TNG) operated by the Fundaci\'on Galileo Galilei of the INAF (Istituto Nazionale di Astrofisica), the William Herschel (WHT) and Isaac Netwton (INT) telescopes operated by the Isaac Newton Group of Telescopes, all of them installed in the Spanish Observatorio del Roque de los Muchachos of the Instituto de Astrof\'{\i}sica de Canarias, in the island of La Palma. 
        Part of the data utilized in this publication were obtained and made available by the The MIT-UH-IRTF Joint Campaign for NEO Reconnaissance. The IRTF is operated by the University of Hawaii under Cooperative Agreement no. NCC 5-538 with the National Aeronautics and Space Administration, Office of Space Science, Planetary Astronomy Program. The MIT component of this work is supported by NASA grant 09-NEOO009-0001, and by the National Science Foundation under Grants Nos. 0506716 and 0907766."
\end{acknowledgements}

\bibliographystyle{aa} 
\bibliography{ACOS2017.bib}

\end{document}